\documentclass[twocolumn,amsmath,amssymb,aps,prb,showpacs]{revtex4}
\usepackage{graphicx}
\usepackage{color}
\usepackage[utf8]{inputenc}

\begin{document}
\title{Magnetic groundstates in a correlated two orbital Hubbard model}

\author{Robert Peters}
\email{peters@theorie.physik.uni-goettingen.de}
\author{Thomas Pruschke}
\affiliation{Department of Physics, University of G\"ottingen, Friedrich-Hund-Platz 1,
37077 G\"ottingen, Germany}

\begin{abstract}
We examine the orbital and magnetic order of the two orbital Hubbard model
within dynamical mean field theory. The model describes the low energy
physics of a partially filled $e_g$-band as can be found in some
transition metal compounds. The model shows
antiferromagnetic as well as ferromagnetic phases. For stabilizing
ferromagnetism we find that Hund's coupling is particularly important.
Quarter filling represents a very special situation in the phase
diagram, where the coupling of spin, charge, and orbital degrees of
freedom are involved.
Exactly at quarter filling we find a
metal insulator transition (MIT) between two almost fully polarized
ferromagnetic 
states. This  MIT can be tuned by changing the local interaction
strength and seems to be a first order transition at zero
temperature. Apart from these ferromagnetic states we were also able to
stabilize antiferromagnetic and charge ordered phases at quarter
filling, depending on the interaction parameters.  
\end{abstract}
\date{\today}
\maketitle
\section{Introduction}
Magnetism is still a very important topic in condensed matter
physics. Analyzing the elements involved in magnetic materials, one
has to conclude that the existence of magnetism is intimately connected to the
presence of partially filled 
d- or f-shells. It is thus not surprising that understanding the influence
of these shells on the low temperature physics is crucial for a proper description of magnetic
materials.

Since d- or f-shells are typically more strongly localized 
than the s- or p-shells of simple metals or semiconductors, they often
are subject to strong correlation effects making their theoretical
study a challenging task.
In this context very interesting materials are the transition metal
oxides \cite{maekawa,imada1998}. One prevalent lattice
structure of transition metal oxides is the cubic perovskite, in which
the transition metal atom sits in the center of an oxygen octahedron.
The states relevant at low temperatures are the d-orbitals of the
transition metal atom. Due to the cubic crystal symmetry the
d-orbitals split into a threefold 
degenerate $t_{2g}$- and a twofold degenerate $e_g$-band, which for the
coordination present in the perovskite structure has the higher energy
compared to the $t_{2g}$-band. Thus not only strong correlations but
also orbital degeneracy plays an important role in the physics of these
compounds. 

To give a specific example for a transition metal oxide, let us
consider the Manganites
\cite{coey1999,salamon2001,tokura2000,dagotto2001,millis1998,tokura2003},
or more precisely La$_{1-x}$Ca$_x$MnO$_3$.
Manganites 
became famous for their colossal magneto 
resistance (CMR). Besides this particular feature they show a very rich
phase diagram with different magnetic and orbitally ordered phases.
In La$_{1-x}$Ca$_x$MnO$_3$ one has to distribute
$4-x$ electrons per site to the d-states of the manganese according to
Hund's rules. Thus, the
electronic configuration in this compound can be modeled by a 
partially filled $e_g$-band close to quarter filling and a half filled
$t_{2g}$-band 
which couples via Hund's coupling ferromagnetically to the
$e_g$-electrons. The hopping  
between the $t_{2g}$-states is very small and thus the $t_{2g}$-states
are often modeled as localized $S=3/2$ spins.
Besides this
electronic part the lattice degrees of freedom and especially Jahn-Teller
distortions are important to correctly describe the physics of Manganites 
\cite{millis1995}.

The ultimate goal surely is to theoretically describe the properties
of materials like  La$_{1-x}$Ca$_x$MnO$_3$ including all the degrees of
freedom mentioned before. However, at least equally important for a proper
microscopic understanding of the physics is to disentangle the contributions
of the different degrees of freedom and identify the individual influence,
in particular to what extent a certain degree of freedom is responsible 
for effects or just follows the lead \cite{Vollhardt:2008}.
For this reason we now leave the special topic of Manganites and focus on the
role of the electronic degrees of freedom, in particular 
the role of interactions on the magnetic and orbital properties of a degenerate 
$e_g$-band. We thus
ignore effects of the $t_{2g}$-spin and the lattice in the following. 
Based on such an investigation, one can later include
further features, like different band widths of the $e_g$-band \cite{koga2004,koga2005,arita2005,costi2007,jakobi2009}, or
additional degrees of freedom,
for example the $t_{2g}$-spin (or band) respectively the strong coupling to
the lattice,  step by step, thereby properly identifying for which
particular features they are actually responsible.
As we will discuss, 
already the simplified situation with only an $e_g$-band present
shows very complex ground state
properties, involving the coupling of charge, spin and orbital degrees
of freedom. 

The article is organized as follows: After this introduction we
establish the model and discuss the methods we have used for solving it. 
In the results section we first analyze the situation for a band
filling between quarter filling and half filling. In the second part
of the results section we particularly address quarter filling
representing a very special point in the phase diagram, at which
different ordered phases compete with each other.
\section{Model and Method}
We model the $e_g$-band as a two orbital Hubbard model
\cite{hubbard1963,hubbard1964,kanamori1963,gutzwiller1963,oles1983}
%, which reads 
\begin{eqnarray}
H&=&\sum_{ij,\sigma}\sum_{m=1}^2t_{ij}c_{i,\sigma,m}^\dagger
c_{j,\sigma,m}-\mu\sum_{i,\sigma}\sum_{m=1}^2n_{i,\sigma,m}\nonumber\\
&&+U\sum_i\sum_{m=1}^2n_{i,\uparrow,m}n_{i,\downarrow,m}-2J\sum_i\vec{S}_{i,1}\vec{S}_{i,2}\nonumber\\
&&+(U^\prime-J/2)\sum_i
(n_{i,\uparrow,1}+n_{i,\downarrow,1})(n_{i,\uparrow,2}+n_{i,\downarrow,2})
\nonumber
\end{eqnarray}
Here $i,j$ label the lattice sites; $m=1,2$ is the orbital and 
$\sigma$ the spin index. Thus, $c_{i,\sigma,m}^\dagger$ creates an
electron at site $i$ in orbital $m$ with spin $\sigma$. As usual, $\mu$
represents the chemical potential with $n=c^\dagger c$ being the
density operator. Note that, in the spirit of the philosophy discussed in the
introduction, we assume the same band structure and -width for
both orbitals in the $e_g$-band and also do not include orbitally off-diagonal
hopping. 

The two particle interaction is parametrized as in  
%[A. Ole\'s]
Ref.~\onlinecite{oles1983},
but we neglect the two particle hopping term
$Jc_{i,\uparrow,m}^\dagger c_{i,\downarrow,m}^\dagger
c_{i,\downarrow,n}c_{i,\uparrow,n}$ \cite{pruschke2005}. This is done
mainly for
numerical reasons, as otherwise one cannot introduce a conserved
orbital quantum number. This of course introduces
an additional approximation; however,
previous studies indicated that this particular term is of minor importance,
at least for ferromagnetic Hund's exchange,
while the inclusion of the rotationally invariant spin exchange appears to
be crucial \cite{pruschke2005}.
Therefore, in our 
model the two particle interaction is given as a local
intra-band interaction with amplitude $U$, a local inter-band
interaction 
with amplitude $U^\prime-J/2$ and a Hund's coupling with amplitude $2J$ between the
spins $\vec{S_1}$ and $\vec{S_2}$  of the two orbitals. 

In order to investigate the possible magnetically, orbitally and charge-ordered
phases of the two orbital model
we use the  dynamical mean field theory (DMFT) 
\cite{pruschke1995,georges1996}. As this mean field theory properly includes
the local dynamics due to electronic correlations, it accounts for
such subtle effects like crossover between itinerant and localized order
\cite{pruschke2005b} and strong reduction of transition temperatures.
As it is well-known, the lattice structure enters the DMFT calculation
only via the  
non-interacting density of states (DOS).
Since we are interested in general
qualitative aspects of ordering phenomena in the two band model and not in the
description of a particular material, we can use this property and
choose a numerically convenient form of the DOS. As has been discussed
extensively\cite{georges1996}, the semi-elliptic DOS obtained from a
Bethe lattice with infinite coordination number
is indeed a numerically convenient choice, which also allows to address the
ordered structures we are interested in, namely
antiferromagnetic N\'eel, homogeneous ferromagnetic, as well as charge and orbital
order. More complex structures, like for example ferromagnetic order in one direction and antiferromagnetic in
another \cite{zoelfl2000} are excluded deliberately.

The DMFT self-consistency maps the lattice onto an effective impurity Anderson
model, which in the present case
becomes a two orbital Anderson model. We solve this Anderson
model with the Numerical Renormalization Group (NRG)
\cite{wilson1975,bulla2008}, which allows us to calculate properties
for wide parameter regions at $T=0$. To reliably calculate spectral
function with 
NRG we use the complete Fock space method
\cite{peters2006,weichselbaum2007}. It must be noted that 
the two orbital model is an extreme case for calculating spectral
functions within the NRG. The calculation of one spectral function requires
approximately $15$GB shared memory and several CPU-hours. As
discretization parameter within the NRG we used $\Lambda=2$, keeping up to
$5000$ states per NRG step in a typical calculation.

\section{Results}
\subsection{Filling range $1<\langle n\rangle\le 2$}
Let us begin the discussion of our numerical results by an overview of the 
magnetic phase diagram as function of filling for $1<\langle n\rangle\le 2$.
The calculations were carried out using a fixed local intra-orbital
interaction $U/W=4$. Here, $W$ is the bandwidth of the semi-elliptic
DOS, which will be used as energy scale throughout this article. This
strength of the interaction is a good guess for transition metal oxides.
As in the one orbital Hubbard model \cite{pruschke2003,peters2009}, 
we also observe that for strong local interaction and  half filling 
\begin{figure}[htb]
\begin{center}
\includegraphics[width=0.48\textwidth,clip]{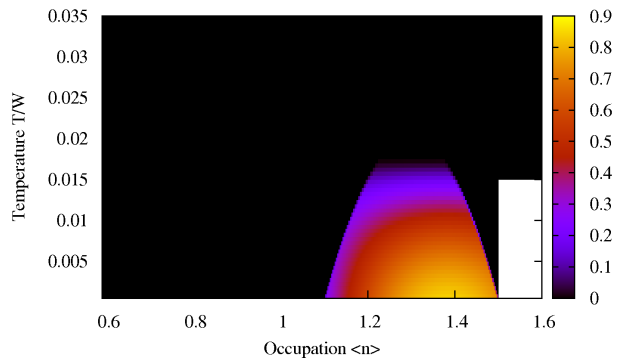}
\end{center}
\caption[]{(color online) Ferromagnetic polarization for
  $J=W$ as function of filling at $U=4W$, $U'=U-2J$ and $T=0$. For fillings
  $n>1.5$ we observe signs of an incommensurate spin density wave
  extending to half filling. The phase diagram was created by fitting
  a smooth surface through approximately $40$ inhomogeneously distributed
  data points. Therefore the location of the phase boundaries
  are only meant as rough sketches.
} 
\label{ferro824}
\end{figure}
the physics is dominated by an antiferromagnetic insulator originating
from super exchange 
\cite{anderson1950}. The N\'eel temperature does not depend on
$J$ or $U^\prime$ within the temperature resolution given by the NRG,
as long as $U$ is the dominating interaction. The 
antiferromagnetic phase can be doped resulting in an incommensurate
spin density wave away from half filling, which can extend to occupations
$\langle n\rangle \approx 1.5$. This is again the behavior expected for
strong local interaction \cite{peters2009}. Note that by virtue of the
self-consistency only phases commensurate with the lattice structure
can be stabilized by a DMFT calculation, but not a truly
incommensurate phase.
Instead, we rather observe oscillations in the magnetization and
occupation during the DMFT self-consistency cycle. Together with
evidence from the one orbital model we can conclude that these oscillations,
appearing when doping the N\'eel state at half filling, are indeed signs of an
incommensurate spin density wave \cite{freericks1995,fleck1999,peters2007}.
For occupations of $\langle n\rangle \lesssim 1.5$ the
physical situation starts to become influenced by the 
double exchange mechanism \cite{zener1951a,zener1951b} and 
we thus expect to find
ferromagnetic order
\cite{fresard1997,held1998,momoi1998,held2000,didkuh2002,fresard2005,sakai2007,kubo2009}.  
\begin{figure}[htb]
\begin{center}
\includegraphics[width=0.48\textwidth,clip]{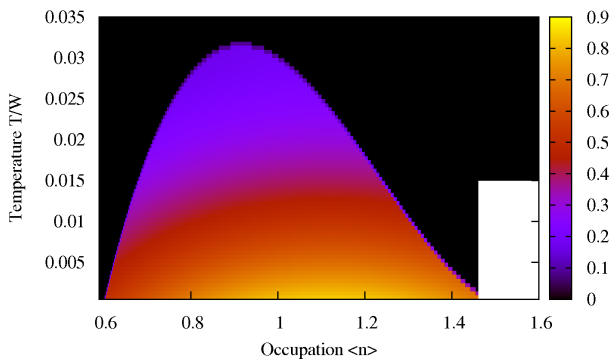}
\end{center}
\caption[]{(color online) The same as in figure \ref{ferro824} for $J=1.5W$.
Compared to $J=W$ the ferromagnetic phase extends to higher
temperatures and larger hole doping.
}\label{ferro832}
\end{figure}
Figure \ref{ferro824} and \ref{ferro832} display the
ferromagnetic  
polarization versus the occupation for two values of the Hund's coupling $J$ for
fixed $U=4W$ and $U'=U-2J$. One can see how
the ferromagnetic phase is stabilized at fillings $n\lesssim 1.5$. 
The blank rectangles close to half filling
in figures \ref{ferro824} and \ref{ferro832} represent the already mentioned
parameter regimes where we found no convergent solution to the DMFT, but an
oscillatory behavior we interpreted as spin density waves. 
As they cannot really be stabilized, the true phase boundaries
cannot be determined in this parameter region.

As
the ferromagnetic double exchange is mostly due to the Hund's coupling,
it is not surprising that for increasing Hund's coupling the ferromagnetic
state is stabilized up to higher temperatures and larger doping and
can even extend beyond quarter filling.  
However, one must emphasize that Hund's coupling alone is not sufficient to enforce
ferromagnetic order in the two orbital model. We furthermore observe
that one in addition needs a rather 
strong interaction parameter $U$ to
stabilize extended regions of ferromagnetism. For example, for $U/W=2$ we
found no ferromagnetism for $1<n<2$. 
\subsection{Magnetic phases at quarter filling, $\langle n\rangle=1$}
\begin{figure}[tb]
\begin{center}
\includegraphics[width=0.45\textwidth,clip]{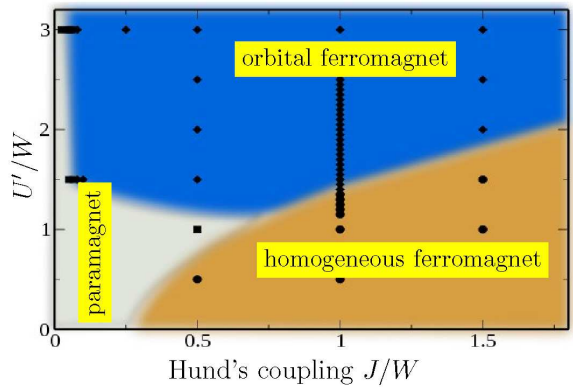}
\end{center}
\caption[]{(color online) Ferromagnetic order at quarter
  filling for $T=0$ and $U=4W$ as function of $U'$ and $J$. Tuning Hund's
  coupling and the inter-orbital density-density interaction, one can observe
  a transition between an orbitally ordered ferromagnetic
  insulator and orbitally degenerate ferromagnetic metal. The symbols
  denote the parameters for which calculations were actually performed.
  The phase boundaries are fits to the calculated points and meant as
guide to the eye.} \label{orbital}\end{figure}

\begin{figure}[htb]
\begin{center}
\includegraphics[width=0.45\textwidth,clip]{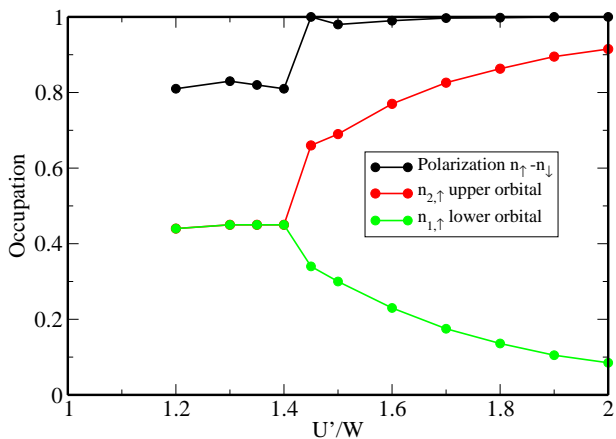}
\end{center}
\caption[]{
(color online) Orbital occupation $\langle n_{m,\uparrow}\rangle $ ($m$
denotes the orbital) and spin polarization for $U/W=4$, $J/W=1$,
$T=0$, and different interaction strengths $U^\prime$. At
$U^\prime/W\approx1.4$ the transition between the homogeneous ferromagnetic
state ($U^\prime/W<1.4$) to the orbital ordered ferromagnet
($U^\prime/W>1.4$) occurs. Lines are meant as guide to the
eye.}
\label{orbtrans}
\end{figure}

\begin{figure}[htb]
\begin{center}
\includegraphics[width=0.45\textwidth,clip]{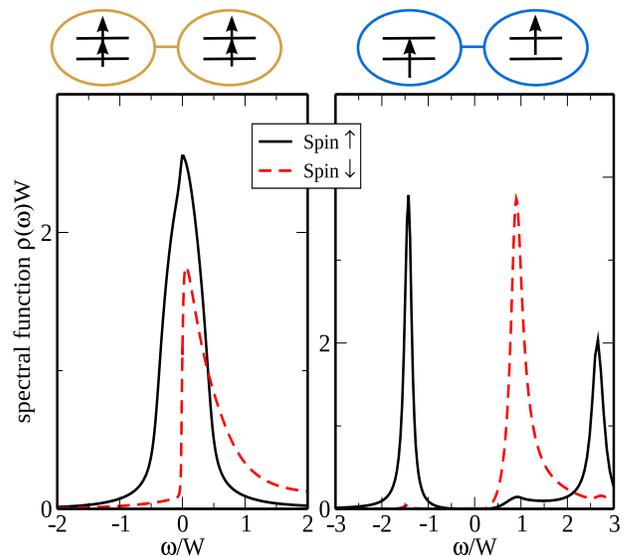}
\end{center}
\caption[]{
(color online) Spectral function and sketch of the electronic configuration.
Left panel: metallic ferromagnetic phase for $U/W=4$, $J/W=1.5$ and
$U^\prime=U-2J$. 
Right panel: insulating ferromagnetic phase for $U/W=4$ and
$J/W=0.5$. $\omega=0$ represents the Fermi energy.}
\label{config}
\end{figure}

Quarter filling, like half filling, represents a very
special point for the two orbital Hubbard model. In a classical
picture there is one electron per site, which can choose between
two orbitals. This picture makes already clear that orbital degeneracy
and fluctuations will play an important role at quarter filling. In
Fig.\ \ref{orbital} 
we show the ferromagnetic ground state phase diagram for $n=1$ as function
of $U'$ and $J$. As noted before, for strong Hund's coupling $J$ and
moderate $U^\prime$
we obtain the orbitally homogeneous ferromagnetic phase discussed in
figure \ref{ferro832}. However, for large enough repulsive inter-orbital
density-density interaction $U^\prime$, we observe that the ferromagnetic
spin alignment is accompanied by an antiferro-orbital order of the conventional
N\'eel-type \cite{kubo2009,roth1966,kugel1973,held1998}.
There is a first order transition between the orbitally
ordered ferromagnetic state and the homogeneous one: both the magnetic
and orbital polarization show jumps when crossing the phase boundary,
see figure \ref{orbtrans}.
Note that  both
states are strongly polarized, i.e.\ the jump in the magnetization
is comparatively small. A more important aspect is that the orbitally
ordered ferromagnetic phase is an 
insulator, while the homogeneous one is a metal. 
This particular difference becomes apparent from
Fig.\ \ref{config}, which shows the spectral function of both ferromagnetic
states at quarter filling. The left (right) panel illustrates the spin and orbital
configuration of the metallic (insulating) phase in the upper part and
the spectral function corresponding to that phase in the lower part.
The Fermi energy lies at $\omega=0$. As can be easily seen,
the homogeneous ferromagnetic state has a large spectral weight at the
Fermi energy, thus representing a metallic system. On the other hand, the
orbitally ordered ferromagnetic state has a gap around the Fermi energy,
thus representing an insulating state. The gap width of the insulating
state decreases when approaching the transition line.
Thus, by varying Hund's
coupling $J$ or the inter-orbital interaction $U^\prime$ one can observe a metal
insulator transition (MIT) between two almost fully polarized
ferromagnetic phases. 
Note that this metal insulator transition is very different from the
usual paramagnetic one,  which appears in the Hubbard model at half filling
as function of $U$. 
The MIT observed here between the ferromagnetic phases
is rather due to a strong inter-orbital density-density interaction, which is
responsible for driving the orbital ordering. As the latter introduces
a doubling of the unit cell, the insulating solution is thus akin to
the antiferromagnetic insulator.
The usual intra-orbital Hubbard interaction $U$ plays a minor role in
this transition. 

It is worth noting that within a simplified two-site
model representing the AB structure of the N\'eel-state the
antiferro-orbital situation wins over the orbitally homogeneous for all
values of $U'$ and $J$ due to virtual hopping and the gain of Hund's
exchange energy. It is thus the presence of the lattice which allows
for smaller $U'$ through an additional gain in kinetic energy the
formation of the homogeneous metallic ferromagnet. In view of effects
like colossal magneto resistivity it is an obviously interesting
question, what influence external control parameters like temperature
and magnetic field will have when one is close to the phase
transition. These questions are presently under investigation. 

\begin{figure}[tb]
\begin{center}
\includegraphics[width=0.45\textwidth,clip]{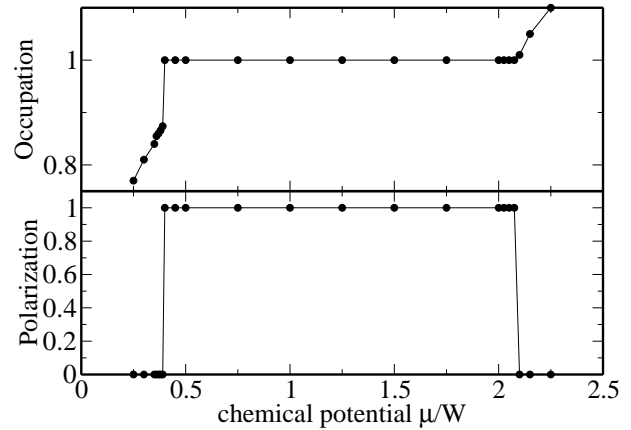}
\end{center}
\caption[]{Orbital ordered ferromagnetic insulator for
  $U=4W$, $J=W/2$, $U^\prime=U-2J$ as function of the chemical
  potential. The upper (lower) panel shows the occupation
  (polarization). Notice the jump in these
  quantities.} \label{orbdop}\end{figure}
Another important aspect is the dependence of the phases upon doping. 
For the homogeneous ferromagnetic phase we already saw when discussing 
figure \ref{ferro832}, that its filling can be varied smoothly. 
The dependency on the chemical potential
of the orbital ordered ferromagnet, on the other hand, is dramatically
different and can be seen in figure
\ref{orbdop}. The plot shows that for a critical chemical potential the
filling of the system jumps from approximately $n=0.88$ to quarter
filling $n=1$. At the same critical chemical potential also the polarization
jumps from a non-polarized phase to nearly fully polarized. In other
words, there is a first order transition between a paramagnetic phase
with filling less than one and an orbitally ordered ferromagnetic state
at quarter filling. Consequently  the electronic system shows 
phase separation between these states and the precise physics will
depend on additional interactions, like long-range Coulomb
interaction, or additional degrees of freedom like the lattice. 
As already noted before, this appearance of phase separation is similar to
what we find in the Hubbard model in the antiferromagnetic phase
at half filling\cite{dongen1994,dongen1995,zitzler2002,peters2009},
and thus seems to be a 
generic feature of the symmetry 
broken phases on an AB lattice. Note, however, that longer-ranged hopping can
actually destroy this phase separation\cite{peters2009}, depending on
the sign of  
the additional hopping and the type of doping.

\begin{figure}[htb]
\begin{center}
\includegraphics[width=0.45\textwidth,clip]{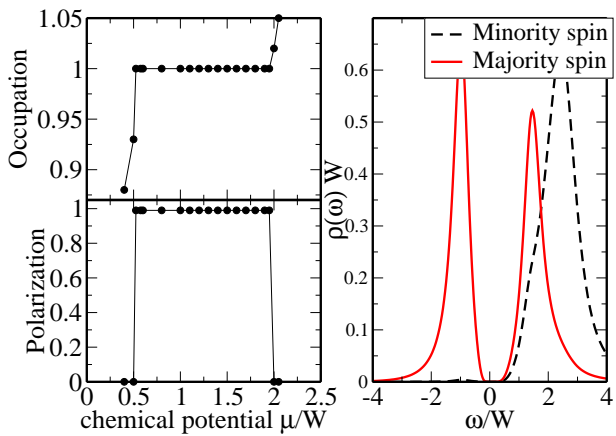}
\end{center}
\caption[]{(color online) Antiferromagnetic insulator for
  $U=4W$, $J=W/2$, $U^\prime=U-2J$ as function of the chemical
  potential. The upper (lower) panel shows the occupation
  (polarization). Notice the jump in the
  properties. The right panel shows the spectral
  function.}
\label{antquart}
\end{figure}

In the region of large $U'$, where the orbitally ordered ferromagnet is found,
we were also able to stabilize an
antiferromagnetic phase at quarter filling. As the system at quarter filling is
dominated by ferromagnetic double exchange, one
actually does not 
expect such a phase here.
Like the orbitally ordered ferromagnet, this antiferromagnetic phase also exists
only exactly at quarter filling. The spectral function and the doping
dependence can be seen in 
figure \ref{antquart}. The spectral function shows that also this
state is a perfect insulator. When trying to dope it we again find
phase separation to a paramagnetic metal away from quarter filling. In
order to find out which state is the thermodynamically stable one, we
calculated the energy of both states.
The result is
that, as expected, the orbital ferromagnet has the lower energy, thus
is the thermodynamically stable one. Looking at the different terms in
the energy, the kinetic 
energy gives a larger decrease for the antiferromagnetic state
than for
the ferromagnetic state, while the interaction terms increase the
energy of the antiferromagnet. Varying the parameters $J$ and
$U^\prime$, we always find the antiferromagnetic state having the
higher energy.

\subsection{Charge ordering at quarter Filling}
For large Hund's coupling
$J/2>U^\prime$, the term $U'-J/2$ 
defining the inter-orbital density-density interaction becomes
attractive. Although at first glance  such a large Hund's coupling
appears unphysical, 
one might have situations, for example Jahn-Teller coupling to phonons, which
can lead to additional contributions to $U'$, typically reducing it effectively. In this case such an
attractive interaction 
can effectively be generated and the physics will change
dramatically. The first thing one notes  
is that during a numerical calculation it becomes very difficult to stabilize fillings other than $n=2$ or
$n=0$. Inspired by this difficulty we investigated charge
ordered phases in this parameter regime; and for sufficiently large
Hund's coupling $J$ it is indeed possible to stabilize a charge ordered state
\begin{figure}[htp]
\begin{center}
\includegraphics[width=0.85\linewidth]{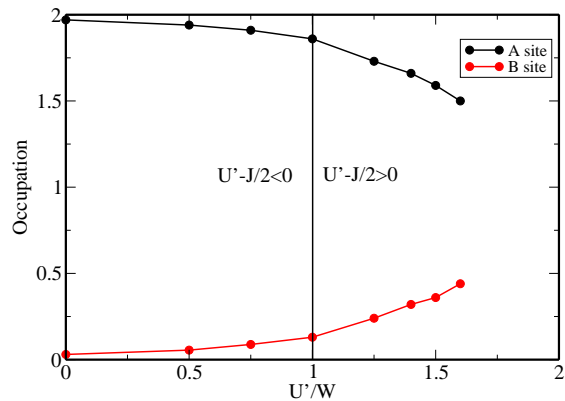}
\end{center}
\caption{Occupation of neighboring sites in the charge ordered state
  for $J/W=2$. The inter-orbital interaction becomes repulsive for
  $U^\prime/W>1$. The lines are meant as guide to the eye.}\label{twocharge1}  
\end{figure}
with alternating almost doubly occupied and nearly
empty sites and an average occupation of $n=1$, see figure \ref{twocharge1}. 
Interestingly, this state
seems to reach into the regime with $U^\prime\gtrsim J/2$, already representing
a repulsive inter-orbital interaction. If $U^\prime$ is increased further, this
charge ordered state finally becomes unstable for $U'_c/W\approx1.7$,
i.e.\ we obtain a quarter filled
paramagnetic state. Note that we do not observe a vanishing order parameter
but rather a jump as $U'\nearrow U_c'$, indicating a first order transition here, too.
Another open point is how the charge ordered state connects to the magnetic
phases present in this parameter region. As we have not yet been able
to perform calculations allowing for both charge and magnetic order,  
we cannot tell whether there is a direct
transition into one of the ferromagnetic phases rather than to the
paramagnetic state. For the magnetic properties of the charge ordered
state one
may expect some kind of magnetic order between the half filled sites,
which however requires larger unit cells to be used in the
calculations. 
Besides magnetism, the charge density wave state is a perfect
insulator, too, and  
again it is not possible to dope this state away from quarter
filling. 
\section{Summary}
In conclusion, we have analyzed the magnetic phase diagram of a two
orbital Hubbard model within DMFT and NRG. While around half filling
the system behaves quite similar to the one orbital Hubbard model, there occurs
an extended ferromagnetic phase for occupation $\langle n\rangle< 1.5$. Such a
ferromagnetic phase cannot be observed for the one orbital case on a
Bethe lattice with semi elliptic DOS and is due to double
exchange mechanism present in the two-orbital model. 
With increasing Hund's coupling $J$ this ferromagnetic phase
becomes more and more extended and finally can also be observed for
occupation smaller than 1. 

A particularly important point in the phase
diagram is quarter filling, where we could observe four different
ordered phases, and an especially interesting feature of the quarter-filled case
is the presence of a
metal-insulator transition between two ferromagnetic states.
The transition seems to be of first order and is driven by
the inter-orbital density-density interaction $U^\prime$. For low
$U^\prime$ the ferromagnetic state is homogeneous and metallic. For
large $U^\prime$, an orbital order can be observed in addition to the
ferromagnetic one, which now is accompanied by an insulating behavior as
the orbital order breaks translational symmetry similar to the N\'eel state
at half filling.

Besides these ferromagnetic states, we also could stabilize an
antiferromagnetic insulating and a charge ordered insulating state at
quarter filling. The antiferromagnetic state exists in the same
parameter region as the orbital ordered ferromagnetic state, but 
has a higher energy, i.e.\ will not be the thermodynamically stable one. 
The charge ordered state can be observed for rather large Hund's coupling $J$
respectively reduced inter-orbital density interaction $U'$. The latter 
situation can for
example be realized in the presence of Jahn-Teller phonons, which definitely
play an important role in two-orbital systems with $e_g$ symmetry.
Thus, the inclusion of lattice degrees of freedom is a very important extension
and presently under investigation.
\begin{acknowledgments}
We thank Prof.\ N.\ Kawakami and Prof.\ A.\ Koga for their hospitality
during a visit to Kyoto and for the discussions on the topic discussed in this paper. RP
also thanks GCOE for financial support during his extended stay at Kyoto
university. We also thank the DFG for financial support through project
DFG-PR298/10 and acknowledge computer support
by the
Gesellschaft f\"ur  wissenschaftliche Datenverarbeitung in G\"ottingen (GWDG) and the
Norddeutsche Verbund f\"ur Hoch- und H\"ochstleistungsrechnen (HLRN).
\end{acknowledgments}


\begin{thebibliography}{22}
\bibitem{maekawa}
S. Maekawa, T. Tohyama, S. Barnes, S. Ishihara, W. Koshibae, and
G. Khaliullin,
\newblock "Physics of Transition Metal Oxides", (Springer), (2004)
\bibitem{imada1998}
M. Imada, A. Fujimori, and Y. Tokura,
\newblock {Rev. Mod. Phys.} {\bf 70}, 1039
\newblock (1998)
\bibitem{coey1999}
J. Coey, M. Viret, and S. Molnar,
\newblock {Adv. Phys.} {\bf 48}, 167
\newblock (1999)
\bibitem{salamon2001}
M. Salamon and M. Jaime,
\newblock { Rev. Mod. Phys} {\bf 73}, 583
\newblock (2001)
\bibitem{tokura2000}
Y. Tokura et al.,
\newblock {\it Colossal Magnetoresistive Oxides} {GORDON AND BREACH
  SCIENCE PUBLISHERS}
\newblock (2000)
\bibitem{dagotto2001}
E. Dagotto, T. Hotta, and A. Moreo,
\newblock {Phys. Rep.} {\bf 344}, 1 
\newblock (2001)
\bibitem{millis1998}
A. Millis,
\newblock {Nature} {\bf 392}, 147
\newblock (1998)
\bibitem{tokura2003}
Y. Tokura,
\newblock {Physics Today} {\bf 56(7)}, 50
\newblock (2003)
\bibitem{millis1995}
A. Millis, P. Littlewood, and B. Schraiman,
\newblock {Phys. Rev. Lett.} {\bf 74}, 5144
\newblock (1995)
\bibitem{Vollhardt:2008} This topic has already been addressed for KCuF$_3$ by
I.~Leonov, N.~Binggeli, Dm.~Korotin, V.I.~Anisimov, N.~Stojić, and D.~Vollhardt,
\newblock Phys.\ Rev.\ Lett.\ {\bf 101}, 096405 
\newblock (2008)
\bibitem{koga2004}
A. Koga, N. Kawakami T. Rice, and M. Sigrist
\newblock {Phys. Rev. Lett.} {\bf 92}, 216402
\newblock (2004)
\bibitem{koga2005}
A. Koga, K. Inaba, and N. Kawakami
\newblock {Prog. Theor. Phys. Suppl.} {\bf 160}, 253
\newblock (2005)
\bibitem{arita2005}
R. Arita and K. Held
\newblock {Phys. Rev. B} {\bf 72} 201102(R)
\newblock (2005)
\bibitem{costi2007}
T. Costi and A. Liebsch
\newblock {Phys. Rev. Lett.} {\bf 99} 236404
\newblock (2007)
\bibitem{jakobi2009}
E. Jakobi, N. Blümer, and P. van Dongen
\newblock {Phys. Rev. B} {\bf 80}, 115109
\newblock (2009)
\bibitem{hubbard1963}
J. Hubbard,
\newblock {Proc. R. Soc. A} {\bf 276}, 238
\newblock (1963)
\bibitem{hubbard1964}
J. Hubbard,
\newblock {Proc. R. Soc. A} {\bf 277}, 237
\newblock (1964)
\bibitem{kanamori1963}
J. Kanamori,
\newblock {Prog. Theor. Phys.} {\bf 30}, 275
\newblock (1963)
\bibitem{gutzwiller1963}
M. Gutzwiller,
\newblock {Phys. Rev. Lett.} {\bf 10}, 159
\newblock (1963)
\bibitem{oles1983}
A. Ole\'s,
\newblock {Phys. Rev. B} {\bf 28}, 327
\newblock (1983)
\bibitem{pruschke2005}
T. Pruschke and R. Bulla,
\newblock {Eur. Phys. J. B} {\bf 44}, 217
\newblock (2005)
\bibitem{georges1996}
A. Georges, G. Kotliar, W. Krauth, and M. Rozenberg,
\newblock {Rev. Mod. Phys.} {\bf 68}, 13
\newblock (1996)
\bibitem{pruschke1995}
T. Pruschke, M. Jarrell, and J. Freericks,
\newblock {Adv. in Phys.} {\bf 44}, 187
\newblock (1995)
\bibitem{pruschke2005b}
T. Pruschke,
\newblock {Prog. Theo. Phys. Suppl.} {\bf 160}, 274
\newblock (2005)
\bibitem{zoelfl2000}
T. Pruschke and M.B. Z\"olfl,
\newblock ``Electronic Structure and Ordered Phases in transition
metal oxides''
%: Application of the Dynamical Mean-Field Theory''
\newblock in {Advances in Solid State Physics} {\bf 40}, B. Kramer (Ed.), Vieweg Braunschweig, Germany, p. 251.
\newblock (2001)
\bibitem{wilson1975}
K. Wilson,
\newblock {Rev. Mod. Phys.} {\bf 47}, 773
\newblock (1975)
\bibitem{bulla2008}
R. Bulla, T. Costi, and T. Pruschke,
\newblock {Rev. Mod. Phys.} {\bf 80}, 395
\newblock (2008)
\bibitem{peters2006}
R. Peters, T. Pruschke, and F. Anders,
\newblock {Phys. Rev. B} {\bf 74}, 245114
\newblock (2006)
\bibitem{weichselbaum2007}
A. Weichselbaum and J. von Delft,
\newblock {Phys. Rev. Lett.} {\bf 99}, 076402
\newblock (2007)
\bibitem{pruschke2003}
T. Pruschke, and R. Zitzler
\newblock {J. Phys.: Condens. Matter}, {\bf 15}, 7867
\newblock (2003)
\bibitem{peters2009}
R. Peters, and T Pruschke
\newblock {New J. Phys.}, {\bf 11} 083022
\newblock (2009)
\bibitem{anderson1950}
P. Anderson,
\newblock {Phys. Rev.} {\bf 79} 350,
\newblock (1950)
\bibitem{freericks1995}
J. Freericks and M. Jarrell
\newblock {Phys. Rev. Lett.} {\bf 74} 186
\newblock (1995)
\bibitem{fleck1999}
M. Fleck, and A. Lichtenstein, and A. Ole\'s, and L. Hedin,
\newblock {Phys. Rev. B} {\bf 60} 5224,
\newblock (1999)
\bibitem{peters2007}
R. Peters and T. Pruschke
\newblock {Phys. Rev. B} {\bf 76} 245101
\newblock (2007)
\bibitem{zener1951a}
C. Zener,
\newblock {Phys. Rev.} {\bf 81}, 440
\newblock (1951)
\bibitem{zener1951b}
C. Zener,
\newblock {Phys. Rev.} {\bf 82}, 403
\newblock (1951)
\bibitem{fresard1997}
R. Fresard and G. Kotliar,
\newblock {Phys. Rev. B} {\bf 56}, 12909
\newblock (1997)
\bibitem{held1998}
K. Held and D. Vollhardt,
\newblock {Eur. J. Phys. B} {\bf 5}, 473
\newblock (1998)
\bibitem{momoi1998}
T. Momoi and K. Kubo,
\newblock {Phys. Rev. B} {\bf 58}, R567
\newblock (1998)
\bibitem{held2000}
K. Held and D. Vollhardt,
\newblock {Phys. Rev. Lett.} {\bf 84}, 5168
\newblock (2000)
\bibitem{didkuh2002}
L. Didkuh, V. Hankevych, O. Kramar, and Y. Skorenkyy,
\newblock{Condens. Matter} {\bf 14}, 827
\newblock (2002)
\bibitem{fresard2005}
R. Fresard, M. Raczkowski, and A. Oles,
\newblock {Phys. Stat. Sol. B} {\bf 242}, 370
\newblock (2005)
\bibitem{sakai2007}
S. Sakai, R. Arita, and H. Aoki,
\newblock {Phys. Rev. Lett.} {\bf 99}, 216402
\newblock (2007)
\bibitem{kubo2009}
K. Kubo,
\newblock {Phys. Rev. B} {\bf 79}, 020407(R)
\newblock (2009)
\bibitem{roth1966}
C. Roth,
\newblock {Phys. Rev.} {\bf 149}, 306
\newblock (1966)
\bibitem{kugel1973}
K. Kugel and D. Khomskii,
\newblock {Sov. Phys. JETP} {\bf 37}, 725
\newblock (1973)
\bibitem{dongen1994}
P. van Dongen
\newblock {Phys. Rev. Lett.} {\bf 74}, 182
\newblock (1994)
\bibitem{dongen1995}
P. van Dongen
\newblock {Phys. Rev. B} {\bf 54}, 1584
\newblock (1995)
\bibitem{zitzler2002}
R. Zitzler, T. Pruschke, and R. Bulla
\newblock {Eur. Phys. J. B} {\bf 27}, 473
\newblock (2002)
\end{thebibliography}
\end{document}